\documentstyle[11pt,newpasp,twoside,epsf]{article}
\def\MSUNYR{\rm M_{\odot}\,yr^{-1}}
\def\MSUN{\rm M_{\odot}}

\def\RSUN{\rm R_{\odot}}
\def\Mdot{ \dot{M}}

\def\be {\begin{equation}}
\def\en{\end{equation}}

\def\edcomment#1{\iffalse\marginpar{\raggedright\sl#1\/}\else\relax\fi}
\reversemarginpar

\begin{document}
\title{Models of Accretion Disks around Young Stars }
 \author{Paola D'Alessio}
\affil{Centro de Radioastronom\'\i a y Astrof\'\i sica, Universidad Nacional Aut\'onoma de M\'exico, Ap. Postal 3-72, Morelia, Michoacan, M\'exico.}
\author{Nuria Calvet, Lee Hartmann }
\affil{Smithsonian Astrophysical Observatory, 
60 Garden Street, MS 42, Cambridge, MA 02138}
\author{James Muzerolle }
\affil{Steward Observatory, University of Arizona, 933 N. Cherry
Ave., Tucson, AZ 85721}
\author{Michael Sitko}
\affil{Department of Physics, University of Cincinnati, Cincinnati, OH 45221-0011}

\begin{abstract}
We discuss the importance of accretion in calculating
disk models for young stellar objects. In particular,
we show that a disk inner rim, irradiated by both
the star and the accretion shocks at the stellar
surface, can naturally explain recent observations of DG Tau
with the Keck interferometer. We present models
for two objects, with mass accretion rates differing
by almost two orders of magnitude, to illustrate
the effects of accretion on the overall disk
structure and emission. 
\end{abstract}

\section{Introduction}

Disks around Classical T Tauri Stars (CTTS) are accretion disks.
This means that a fraction of disk material loses energy and
angular momentum, falling into the gravitational potential 
well of the star.
The standard view of the final accretion process in a low mass star 
is that a well organized stellar magnetic field 
truncates and couples to the inner disk at a few stellar radii 
(K\"onigl 1991; Shu et al. 1994). 
Disk material falls onto the star 
along the stellar magnetic field lines in a magnetospheric 
flow, 
releasing most of the accretion luminosity, 
$L_{shock} \sim G M_* \Mdot/R_* (1-R_*/R_i) \sim L_{acc}$, 
at the stellar surface 
in  hot accretion shocks(Calvet \& Gullbring 1998, CG98).
The emission of these shocks produce  a blue/UV continuum excess 
that veils the stellar photospheric absorption lines. Detailed 
measurement of veiling excesses, combined with models
of the hot continuum,
have resulted in a long list of 
mass accretion rate estimates in CTTS (see Calvet, Hartmann \& Strom 2000, and 
references therein). In addition, models of the gas infalling 
through the
magnetosphere have successfully reproduced blue shifted emission 
peaks, blue line asymmetries and red shifted absorption components 
of emission lines in CTTS
(Hartmann, Hewett \& Calvet 1994; Muzerolle, 
Hartmann \& Calvet 1998; Muzerolle, Calvet \& Hartmann 1998, 2001),
providing strong support to the magnetospheric model
of accretion.

The average mass accretion rate for CTT disks in several known associations 
is $\sim 10^{-8} \ \MSUNYR$ (Gullbring et al. 1998; Hartmann et al 1998.)
For this low mass accretion rate, the infrared SED produced by 
the dust in the accretion disk is dominated by reprocessed stellar radiation 
intercepted by the disk curved surface (Kenyon \& Hartmann 1987;
Calvet et al. 2000). 
Thus, why should one  get into the trouble of including accretion 
in a model describing a disk?
We discuss in this contribution why accretion plays a
key role in disk structure and cannot be ignored.

Models of passive disks (i.e., disks heated only by stellar 
irradiation) have been proposed to explain the observed SEDs of 
CTTS and Herbig Ae stars from IR to mm wavelengths 
(Chiang \& Goldreich 1997).
In contrast to power law models, where mass surface density and temperature 
of the disk are given by arbitrary power laws,  
the passive disk models treat in detail the energy balance 
 between  heating by absorbed stellar radiation and radiative cooling 
at each point. 
In recent years, 
powerful radiative transfer techniques have been developed for
passive disk models, which result in  successful accounting for 
several observed features  (see Dullemond, van Zadelhoff \& Natta 2002; Dullemond 2002; Dullemond \& Natta 2003). 
However, even in the most sophisticated versions of these models, 
the disk mass surface density is 
still assumed to be an arbitrary power law of distance,  
$\Sigma = \Sigma_0 (R/R_0)^{-p}$,  
and $\Sigma_0$ and $p$ are 
adjusted to fit a particular observation, with no 
link to the underlying physics of the disk.

On the other hand, in a steady accretion disk 
the mass accretion rate  is related to 
the angular momentum flux produced by a viscous torque. 
As a consequence, the  mass 
surface density is given by $\Sigma \propto \Mdot/\nu_t$, 
where $\nu_t$ is the viscosity coefficient.
It is common to adopt the $\alpha$ prescription 
(Shakura \& Sunyaev 1974), such that 
$\nu_t = \alpha c_s H$, where $c_s$ is the local sound speed, $H$ 
is the local gas scale height and $\alpha$ summarizes the unknown 
origin of the viscosity. Thus, instead of assuming an arbitrary power law  
for $\Sigma(R)$, 
the surface density is a result of the calculations in 
accretion disk models,
given $\Mdot$ (which can be independently quantified from observations), 
the temperature (which is calculated from viscous and irradiation heating)
 and $\alpha$ (which, in principle,  can be  estimated from 
magnetohydrodynamic simulations; however, in general, is taken to be 
 a free constant parameter).

Irradiated accretion disk models show that 
even for $\Mdot \sim 10^{-8} \ \MSUNYR$, 
 viscous dissipation is an important heating mechanism 
close to the mid-plane and close to the central star ($ R < $ 1 AU)
(D'Alessio et al. 1997, 1999, 2001). 
On the other hand,
stellar radiation intercepted  by the disk upper layers 
is the most important heating 
mechanism of the disk atmosphere and the outer disk 
(Kenyon \& Hartmann 1987; Calvet et al. 1991, 1992 =CMPD; 
D'Alessio et al. 1998, 1999, 2001).
The upper layers of the atmosphere are hotter than the interior 
 (CMPD), and this is a characteristic  
also present in the passive models mentioned above.  
These upper optically thin hot layers are  probably responsible for 
some of the observed emission features (silicate 10 $\mu m$ band,  
Natta, Meyer \& Beckwith 2000; CO overtone emission, Najita et al. 1996). 
Their reprocessed radiation heats the disk interior, 
but do not contribute much to the continuum, specially when 
scattering of stellar radiation by atmospheric dust grains is included 
(D'Alessio et al. 1998; Dullemond \& Natta 2003).

In this paper we illustrate the importance 
of taken into account  the accretion nature of the disk. 
First of all, the disk mass surface density is related to its mass accretion
rate and viscosity, and the local viscous dissipation should be included
as a heating source, even if the mass accretion rate is small.
Second, the disk receives radiation from both the star
and the accretion shocks at the stellar surface.
This becomes particularly important for those objects
where the accretion luminosity is comparable to the
stellar luminosity, and has profound effects on the
structure of their inner disk, as discussed below. It is specially relevant
for high resolution interferometric observations,
and opens up a number of possibilities to unveil  the
physics of accretion.

\section{Irradiated Rim}

Muzerolle et al. (2003a, MCHD) have presented near-infrared spectra of 
the excess continuum
emission from the innermost regions of 9 CTTS.
The shape of this excess is consistent 
with that of a single 
temperature black-body with $T \sim 1200-1400$ K, 
which are the expected sublimation temperatures for 
silicates under typical disk conditions (Pollack et al. 1994).
The observed SEDs agree with that of a rim at the dust sublimation radius, 
receiving radiation frontally. A rim irradiated 
by the central star has been 
proposed by Natta et al. (2001), Dullemond, Dominik \& Natta (2001) 
to explain the observed excess at 3 $\mu m$ in the SEDs of Herbig Ae stars, 
and by Tuthill, Monnier \& Danchi (2001) and
 Monnier \& Millan-Gabet (2002) to explain interferometric near-IR 
images  of disks around Herbig Ae and CTT stars. 
However, in the CTTS case, MCHD find
no direct relationship between the radius
of the rim and the stellar luminosity; rather,
this radius increases with accretion luminosity.
This is naturally explained when the fact that the disk is
irradiated by both the stellar and the accretion luminosity
is taking into account (MCHD.)
In the following section we outline 
the model  for the temperature 
at the rim atmosphere, which can explain
the MCHD observations,
and we show 
how the rim radius depends on the accretion and
stellar luminosities as well as on the dust properties.

\subsection{Model for the irradiated Rim of an accretion disk}

We follow a simplified version of CMPD's treatment 
to describe the radial temperature distribution in the rim atmosphere, 
assuming it has plane-parallel geometry and constant opacity. 
The radiation field is separated into two frequency ranges,
one characteristic of the disk (and the related quantities 
have  subindex ``d'') and one characteristic of the incident 
radiation (with subindex ``i''). In the case of the  rim, the incident 
field is given by the superposition of the  
stellar radiation field 
(characterized by the star effective temperature, $T_*$) 
and the accretion shock radiation field (characterized by a temperature $\sim$
8000 K and a luminosity $L_{shock}$, CG98).

For simplicity, we assume that the scattering of incident radiation 
is essentially forward, which is 
equivalent to take the albedo to be equal zero in CMPD equations (i.e., 
$J_s=H_s=0$).
Also, we assume that the incident radiation reaches the 
rim with an angle 0 relative to the normal to the surface, i.e., cos$\theta_0 \, =  \, \mu_0=1$, 
and that there is no other heating source than the incident radiation.
Thus, at every depth into the rim atmosphere the net outward 
radial flux at the 
disk frequency range is equal to the incident stellar flux, i.e,

\be
H_d(\tau_s) = {F_0 \over 4 \pi} e^{-\tau_s}
\en
where $F_0=(L_*+L_{shock})/4 \pi R^2$ and $\tau_s$ is the optical 
depth to the stellar radiation in the incident direction (i.e., 
along rays parallel to the disk mid-plane, perpendicular to a 
cylindrical surface at the dust sublimation radius).

The zeroth moment of the transfer equation can be written as (Mihalas 1978),

\be
{dH_d \over d\tau_d} =  J_d - S_d
\en
where $\tau_d$ is the optical depth at the disk frequency range, 
$S_d$ is the local source function and $J_d$ is the mean intensity. 
CMPD assume strict LTE, i.e., $S_d=B_d=\sigma T_d^4/\pi$. 
Here, we will include a term corresponding to the emissivity by scattering 
at the disk frequency range, i.e., 
$S_d = (\kappa_d B_d + \sigma_d J_d)/(\kappa_d+\sigma_d)$, assuming 
isotropic scattering at this frequency range (Mihalas 1978). 
With this source function, 
the zeroth moment of the transfer equation is given by  

\be
B_d = J_d - {\chi_d \over \kappa_d} {d H_d \over d \tau_d}.
\en

Using the Eddington approximation, the first moment of the transfer 
equation is
\be
{dJ_d \over d \tau_d} = -3 H_d = -3 {F_0 \over 4 \pi} e^{-q \tau_d}
\en
where we have defined $q=\tau_s/\tau_d$. Integrating $J_d$ from 
eq(4), using 
$J_d(0)=2 H_d(0)$  as the boundary condition, and  
substituting $J_d$ and $dH_d/d\tau_d$ from eq(1) 
in eq(3), the temperature as a function of $\tau_d$ 
can be written as

\be
T_d(\tau_d)^4 = {F_0 \over 4 \sigma_R } \biggl [ \biggl (2 + {3\over q}\biggr ) +\biggl (q {\chi_d \over \kappa_d}- {3 \over q} \biggr) 
e^{-q \tau_d}\biggr]. 
\en 

The temperature at the innermost radius of the rim, where $\tau_d= \tau_s=0$, 
should be equal to the silicates sublimation temperature (assuming 
the inner gaseous disk has a negligible radial optical depth). Thus, 
the rim radius $R_{sub}$ is 

\be
R_{sub} = \biggl [ \biggl ( {L_* + L_{shock} \over 16 \pi \sigma_R} 
\biggr ) \biggl ( 2 + q {\chi_d \over \kappa_d} \biggr ) \biggr ]^{1/2} 
{1 \over T_{sub}^2}.
\en

Eq(6) is essentially similar to the equation for the rim radius proposed by  
Tuthill et al. (2001), except for the enhanced  heating due to the accretion 
shocks at the stellar surface (given by $L_{shock}$),
and the addition of 2 to the opacity term inside the
parenthesis, which arises from the explicit inclusion
of the disk radiation field.

The temperature at the level where $\tau_d=2/3$, which characterizes the 
emergent flux, is 

\be
T^4\biggl (\tau_d={2\over3}\biggr )= 
T_{eff}^4= \biggl ( {L_* + L_{shock} \over 16 \pi \sigma_R}\biggr ) 
\biggl ( 2 + {3 \over q} +  \biggl({q \chi_d \over \kappa_d} - {3 \over q} \biggr) e^{-2q/3} \biggr ] .
\en

Modeling of the observed near IR SEDs of CTTS indicates
that the contribution of the optically thin 
layers of the rim (with dust) to the SED is negligible, so that
the emergent intensity $I_\nu$ differs from $B_\nu(T_{eff})$  
by only few percent (MCDH). 
Also, we find that $T_{eff}$ 
is lower than $T(0) = T_{sub}$ within $\sim 100$ K, so that the rim 
atmosphere is almost radially isothermal (MCDH).

Finally, the emergent flux is $F_\nu = I_\nu \Omega(i)$, where $\Omega(i)$ 
is the solid angle subtended by the rim as seen by the observer and 
$i$ is the inclination angle between the disk symmetry axis and the line 
of sight.  The rim height is $z_{rim}$, defined as the height 
where the optical depth to the stellar radiation impinging radially 
becomes unity at the rim radius $R_{sub}$.
For an accretion disk this height can be estimated  
considering the rim to be 
vertically isothermal with a mass surface density 
given by  $\Sigma = {\Mdot/6 \pi \nu_t}[1-(R_*/R_{sub})^{1/2}]$ 
(Muzerolle et al. 2003b).  We find typically $z_{rim} \sim 4 H_{rim}$, 
where $H_{rim}$ is the gas vertical scale height of the rim (assumed 
to be vertically isothermal).

Defining $\delta = z_{rim} \tan i/R_{sub}$, the solid angle of 
the rim is 

\be
\Omega(i)= \pi \biggl ( {R_{sub} \over d} \biggr )^2 \cos i \ \ \ \ \ {\rm for} \ \ \  \delta > 1
\en

\be
\Omega(i)= 4 \biggl ( {R_{sub} \over d} \biggr )^2 {\cos i \over 2}  ( \sin^{-1} \delta  + \delta \sqrt{1 - \delta^2} )  \ \ \ \ \ {\rm for}  \ \ \ \delta < 1
\en
Both expressions multiplied by $B_\nu(T_{eff})$, 
 are consistent with the formulae for the emergent flux 
given in the Appendix B of Dullemond et al. (2001), except 
for our different definition of $R_{sub}$, which assumes that 
the dust is optically thin at the sublimation radius, and 
for using a $T_{eff}$ different from the sublimation temperature.

\section{Results}
In this section, we discuss irradiated accretion disk models,
including the rim, that explain high resolution IR observations or SEDs 
of different objects. 
The rim model just described fits the near IR spectra  
of several  CTTS, with rim radii in the range
0.07-0.54 AU, increasing with higher stellar and accretion luminosities
(MCDH). Here, we first show that this model can 
explain the recent interferometric observations
presented by the Keck team. Then, we present models for two systems with
widely different accretion luminosities, to illustrate the
effects of accretion.

\subsection{DG Tau}

Keck interferometric observations of DG Tau (Colavita et al. 2003) 
showed that the system was resolved at K, and derived
a size of $\sim$ 0.12-0.24 AU. 
This radius is $\sim$ 3 times larger than 
the dust sublimation radius inferred assuming the rim is irradiated only by 
the central star. Specifically,
for an optically thick wall, 
$R_{sub}^\prime = \biggl ( {L_* / 16 \pi \sigma_R T_{sub}^4} \biggr ) ^{1/2}$. 
Using typical parameters for a K7-M0 central star 
($T_*= 4000$ K, $R_*=2 \ \RSUN$ and $M_*=0.3 \ \MSUN$), 
and a silicate sublimation temperature $T_{sub}=1400$ K,   
$R_{sub}^\prime \sim 0.07$ AU.
However, DG Tau is a ``continuum star'', with a mass accretion 
rate $\Mdot = 5 \times 10^{-7} \ \MSUNYR$ (Gullbring et al. 2000). Assuming 
 the typical dust properties derived for the 
sample of CTTS studied by MCHD
(i.e., $q=1$, $\chi_d/\kappa_d=2$), we find that 
$R_{sub} \sim 0.2$ AU, which is $\sim 3$ time larger than $R_{sub}^\prime$, 
and consistent with the observed inner radius of DG Tau. 

\subsection{LkCa 15}
We have calculated a model for the structure and emission of LkCa 15, 
which is a low mass accretion rate CTTS. The details of the calculation 
of an irradiated 
accretion disk model are described by D'Alessio et al. (1998, 1999, 2001).
The input parameters constrained by observations 
are $\Mdot= 1.3 \times 10^{-9} \ \MSUNYR$,  
$T_*=4350$ K, $R_*=1.5 \ \RSUN$, $M_*=0.8 \ \MSUN$ (Hartmann et al. 1998)
Figure 1-right panel shows the comparison between the observed and predicted SEDs. 
We explain the shape of the observed SED 
between 60 $\mu m$ to 1.3 mm, 
assuming that the outer disk ($R >$ 3 AU) has a uniform 
grain size distribution 
characterized by $n(a) \propto a^{-3.5}$, with $a_{max} =$ 1 mm. 
These are typical dust properties such that an irradiated disk model 
with the median $\Mdot$ and a typical central star, 
fits the median SED of CTTS in Taurus ( D'Alessio et al. 2001).  
We adopt a maximum disk radius $R_d = $ 200 AU and $\alpha=0.001$, 
corresponding to a disk mass $\sim 0.04 \ \MSUN$. 
Qi et al. (2003) find that the mm SED does not show a single spectral index, 
which is naturally explained, by taking into account
the contribution of optically thick regions at small radii.

In this model, the disk rim is placed 
at $R_{sub} \approx $ 0.06 AU. Being $L_* > L_{shock}$, the rim is 
mostly heated by stellar radiation.
There is a shadowed 
region between the rim 
and $\sim$ 0.16 AU, which produces a negligible contribution to the SED.
We assume that this region  is only being heated by viscous dissipation.
Finally, there is an intermediate region between $\sim$ 0.16 and 3 AU, 
where the disk atmosphere is characterized by smaller grains than 
the disk interior and the disk outer regions. 
This region is responsible for the observed 10 $\mu m$ silicate band
which is formed in the upper layer, heated by direct 
stellar radiation (Natta  et al. 2000). 
In order to fit the shape of the silicate band, we have used
optical constants for olivine, MgFe SiO$_4$ (J\"ager et al. 1998).
To produce a continuum around 10 $\mu m$ consistent 
with the observed value,  the dust to gas
mass ratio of the upper atmosphere should be a factor $\sim 0.05$  smaller
that the standard value, probably reflecting that dust growth and 
settling has taken place in the inner disk.
In principle, the details of this model can be tested by high 
resolution observations capable of resolving scales 
between 0.4 and 20 mas.
 
\subsection{DR Tau}

DR Tau is a continuum star, characterized by a strong UV/optical continuum.
We have calculated a model for the disk of DR Tau, with 
the following parameters  (MCHD)
 $T_*=4000$ K, $R_*=1.9 \ \RSUN$ and $M_*=0.9 \ \MSUN$ for the central star.
We take $\Mdot=5 \times 10^{-7} \ \MSUNYR$, 
roughly consistent with the value inferred from the optical/UV excess 
(Kenyon et al. 1994; Gullbring et al. 2000). A uniform maximum grain size 
$a_{max}=$ 1 mm, $\alpha= 0.01$, $R_d=$ 50 AU, 
corresponding to  a disk mass $M_d \sim 0.2 \ \MSUN$.
In contrast to the case of LkCa 15, accretion luminosity
determines the structure and emission of the entire disk for this object.
The rim radius is $R_{sub} \, \sim \, 0.15$ AU,  and it is heated 
mostly by the hot continuum produced in the accretion shocks at the 
stellar surface. Figure 1-left panel shows that 
the rim emission dominates the SED between 3 and 6 $\mu m$. 
Local viscous dissipation is 
important in heating the disk photosphere for $R \la $ 3 AU and 
heating the mid-plane at every radius ($ R \la$ 50 AU).
However, the photosphere of the outer disk is heated mostly by radiation from 
the accretion shocks. 
This model can also be tested by high resolution observations.

\begin{figure}
\plotone{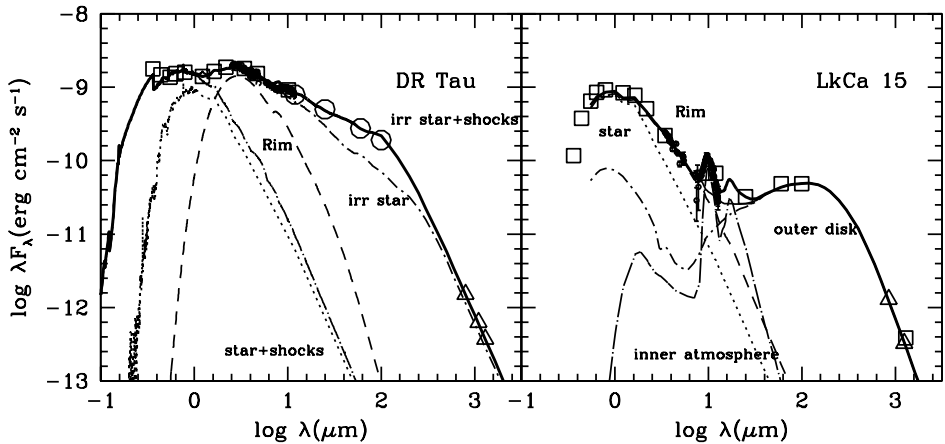}
\vskip-6cm
\caption{Right panel:SED of LkCa15 showing different contributions.
Left panel: SED of DR Tau showing the contributions 
of different zones. For comparison, also a 
SED of the same disk model only irradiated by 
the central star is shown. 
See D'Alessio et al. (2003) for references
to observational data.
}
\end{figure}

\section{Conclusions}

We have shown that accretion plays an important
role in determining the structure and emission of
disks around CTTS and cannot be dismissed. The
surface density of the disk depends on the
mass accretion rate, and moreover,
accretion energy determines the location
of the transition between dust and gas in the inner
disks, with direct implications for interferometric
observations.
In particular, by including the contribution
of accretion, the puzzling Keck interferometer
observations of DG Tau, with an disk inner radius larger than the 
sublimation radius expected from stellar luminosity
alone, can be understood.

\acknowledgements
PD would like to thank the organizing Comittee of IAU Symposium 221 for
a travel grant. MS acknowledges a grant from NASA Origins of
Solar Systems NAG5-9475.
 Support for this research comes from NASA Origins of
Solar Systems grant NAG 5-9670  and grants from UNAM-DGAPA/PAPIIT 
and Conacyt, M\'exico.

\end{document}